\documentclass[pdflatex,sn-mathphys-num]{sn-jnl}

\usepackage{graphicx}%
\usepackage{multirow}%
\usepackage{amsmath,amssymb,amsfonts}%
\usepackage{amsthm}%
\usepackage{mathrsfs}%
\usepackage[title]{appendix}%
\usepackage{xcolor}%
\usepackage{textcomp}%
\usepackage{manyfoot}%
\usepackage{booktabs}%
\usepackage{algorithm}%
\usepackage{algorithmicx}%
\usepackage{algpseudocode}%
\usepackage{listings}%
\usepackage{booktabs}
\usepackage{tabularx}
\usepackage{fancyhdr}
\usepackage{hyperref}
\usepackage{float}

\theoremstyle{thmstyleone}%
\theoremstyle{thmstyletwo}%

\theoremstyle{thmstylethree}%
\begin{document}

\title[Article Title]{Rhetorical-Role-Aware Retrieval-Augmented Generation for Legal Question Answering over Indian Supreme Court Judgments}


\author*[1]{\fnm{Sayed Ayaan} \sur{Ahmed Sha}}\email{sayedayaan810@gmail.com}

\author[1]{\fnm{Sangeetha} \sur{Sivanesan}}

\author[2]{\fnm{Anand Kumar} \sur{Madasamy}}

\author[2]{\fnm{Navya} \sur{Binu}}

\affil[1]{\orgname{National Institute of Technology}, \city{Tiruchirappalli}, \postcode{620015}, \state{Tamil Nadu}, \country{India}}

\affil[2]{\orgname{National Institute of Karnataka}, \city{Surathkal}, \postcode{575025}, \state{Karnataka}, \country{India}}


\abstract{This research paper proposes a Retrieval Augmented Generation (RAG) framework that is specific to the legal field in order to assist interactive retrieval and reason about judgments from the Supreme Court of India. The solution uses an enhanced version of RAG framework which consists of rhetorically based chunking, fusion-based retrieval, and cross encoder reranking methods to increase the relevancy of the information retrieved. In order to improve conversations, the proposed framework uses chat history along with query classification and rewriting in order to understand user intention from successive queries. Additionally, there are features that take into account structural aspects of legal documents, such as isolated names of judges that could have an impact on retrieval quality. The evaluation was done using the DeepEval framework and demonstrated strong performance on metrics including contextual recall and answer relevancy, which proves that the framework is
very effective in dealing with legal question-answering tasks that require a lot of context. The results emphasize the importance of domain specific enhancements in developing legal AI systems that are both reliable and explainable.}




\maketitle

\section{Introduction}\label{sec1}

Large language models (LLMs) are evolving rapidly and have enabled effective analysis and understanding of legal texts. With their strong natural language capabilities, these models have found various applications in a wide range of legal tasks, including question answering, document summarization, statutory interpretation, and intelligent legal assistance. Legal question answering has received considerable attention because it tries to provide correct, contextually relevant responses to queries put by the general public, lawyers and researchers. But due to the complex structure of court documents, the use of legal terminology, and the requirement to identify relevant material scattered throughout several sections of a legal document make it challenging to respond to legal issues. Traditional retrieval approaches often suffer from vocabulary mismatch, while dense retrieval techniques may overlook legally significant terms such as case citations, statutory provisions, and section numbers. Current research has increasingly focused on RAG frameworks being integrated with information retrieval along with large language models ability to reason. Although these methods have yielded promising results, they treat legal texts as homogeneous text collections and extract information based on lexical or semantic similarities. Because of this, they often fail to adapt retrieval techniques to different categories of legal topics and ignore the underlying structure of judicial decisions.\\

Our work proposes a  rhetorical-role-aware retrieval-augmented generation framework for legal question responding over Indian Supreme Court rulings.
The framework utilizes rhetorical role annotations to divide judgments into meaningful segments. It uses a hybrid technique that combines dense and lexical retrieval based on BM25 \citep{robertson2009probabilistic}. This method also employs intent-aware role filtering to improve retrieval accuracy by matching various legal questions with the rhetorical roles most likely to contain relevant information. A cross encoder model is then applied to rerank the chosen passages, which are subsequently fed into a Large Language Model(LLMs) to produce the final response.

\section{Related Work}\label{sec2}
Artificial intelligence has greatly advanced thanks to recent developments in large language models (LLMs), allowing systems to process, interpret, and generate natural language in a manner that closely resembles human communication. These capabilities have attracted significant interest in the legal domain, where the large volume, complex structure, and specialized nature of legal documents often make information retrieval and analysis a challenging task. The effectiveness of large scale language comprehension and transfer learning was shown by the success of pretrained language models like BERT \citep{devlin-etal-2019-bert} and GPT \citep{radford2018improving}. \citep{lewis2020retrieval} introduced RAG, that combines non parametric memory, a dense vector index of Wikipedia accessed via a neural retriever and parametric memory, a pretrained seq2seq model like BART. \citep{arora2025building} created an LLaMA \citep{touvron2023llama} based legal chatbot that creates context-aware responses for non-expert users and uses natural language interactions to deliver legal information. LegalEase, a RAG based legal QA system that combines semantic retrieval with large language models was created by \citep{mubeen2025redefining} to generate grounded answers as in Bharatiya Nyaya Sanhita (BNS) and the Indian Penal Code (IPC). Another study by \citep{nigam2023legal} highlighted that GPT-3 (Davinci) combined with Ada or Instructor-XL embeddings achieved the best results on evaluation of Indian criminal legal QA. \citep{wahidur2025legal} proposed the Legal Query RAG (LQ-RAG) that addressed the common limitations of conventional RAG systems, such as hallucinations and inaccurate responses. The framework employed a recursive feedback mechanism and a dedicated evaluation agent to verify retrieved evidence and refine generated answers  \citep{nigam2025nyayarag} proposed \textbf{Nyayarag}, a a retrieval-enhanced framework for generating explanations and forecasting court judgements in the Indian legal system. The findings of the experiment showed that adding outside legal information improves the quality of explanations and predictions. \textbf{TaxFlow}, a hybrid RAG framework for statutory question answering system, combines dense retrieval using BGE embeddings with sparse retrieval based on BM25/SPLADE, followed by cross-encoder reranking and LLaMA-3-based answer generation \citep{karna2026hybrid}. \citep{devaraj2023development} created a chatbot for legal documents that uses a Large Language Model and semantic retrieval to respond to user queries about uploaded legal documents via a conversational interface. \citep{louis2024interpretable} presented LLeQA dataset and an end-to-end retrieval-then-read framework that combines an instruction-tuned large language model with a lightweight bi-encoder retriever to produce thorough responses from relevant legal sections.\\ \citep{askari2024answer} proposed $CE_{FS}$, a cross-encoder-based framework that enhances legal answer retrieval by incorporating fine-grained structured information during the reranking process. The authors also introduced the LegalQA benchmark dataset and demonstrated that $CE_{FS}$ consistently outperformed existing cross-encoder reranking models Based on the U.S. federal case law, \citep{hou2025clerc} developed \textbf{CLERC}, a comprehensive benchmark for retrieving legal knowledge and retrieval-augmented creation. The dataset offers an extensive resource for creating and assessing RAG and long-context legal retrieval systems.
\citep{pipitone2408legalbench} introduced \textbf{LegalBench-RAG}, benchmark for assessing retrieval performance in legal RAG systems. Based on Mistral-7B, SaulLM-7B \citep{colombo2403saullm} has been tailored to the legal area. The model achieved SOTA performance on tasks involving the understanding and creation of legal texts by incorporating legal instruction customization. Semantic segmentation \citep{malik2022semantic} was addressed by categorizing sentences into specific rhetorical roles, such as Arguments, Facts, Ratio Of The Decision (ROD), Argument or Ruling By Lower Court (RLC). \citep{paul2023pre} introduced InLegabBERT, a pretrained model for Indian legal texts, showing that legal-domain pre-training greatly enhances performance on tasks like court judgement prediction, legal statute identification, and semantic segmentation. \citep{abdallah2023exploring} did a comprehensive examination on benchmark datasets, deep learning models, and recent developments in the field to provide a thorough overview of legal question answering. The survey also suggested a taxonomy of legal QA systems along with the need for more efficient retrieval and reasoning methods to enhance legal question answering.

\section{Methodology}\label{sec3}

\begin{figure}[h]
  \centering
  \includegraphics[width=0.85\linewidth]{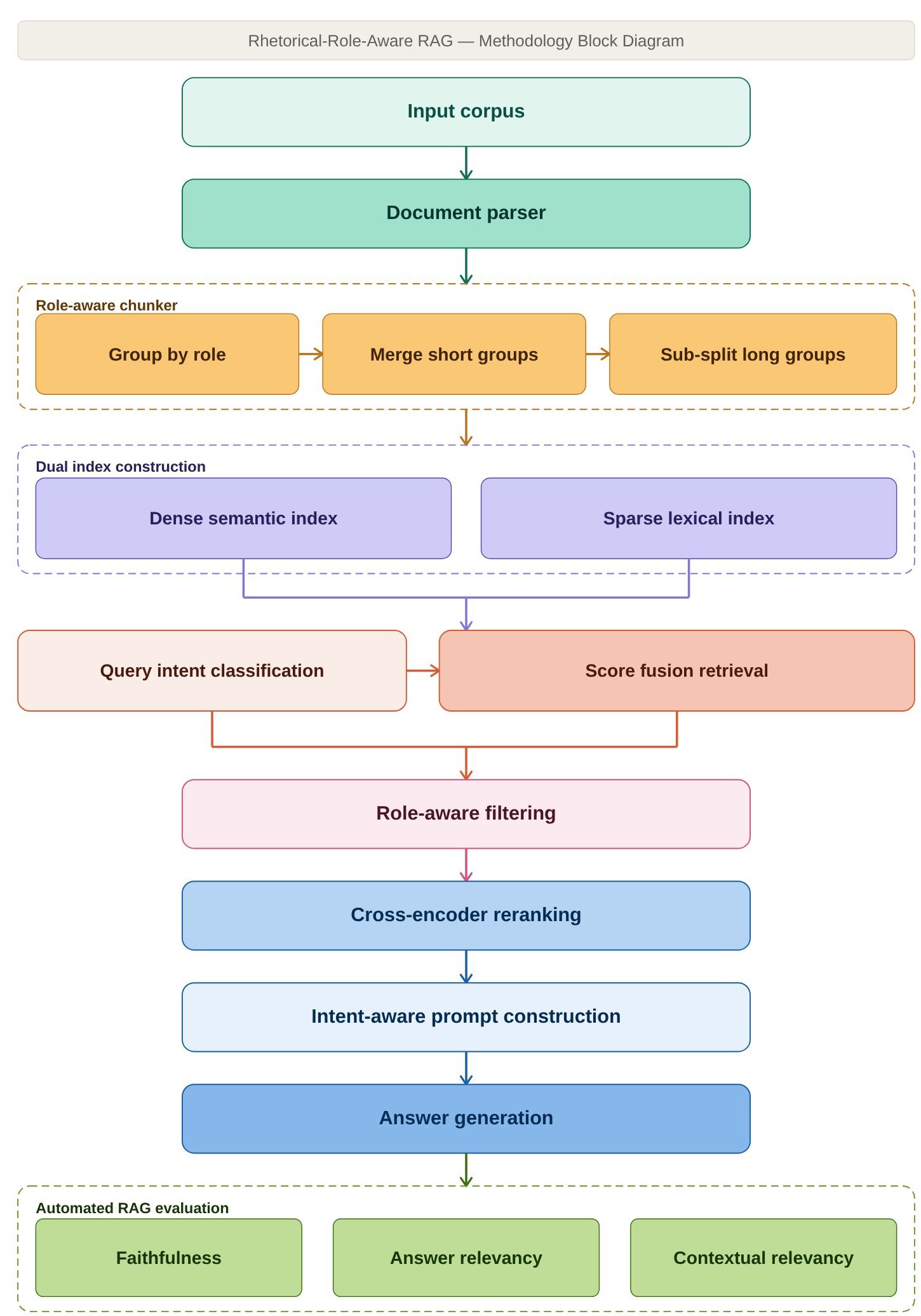}
  \caption{Rhetorical-role-aware RAG system.}
  \label{fig:methodology}
\end{figure}

Our approach presents a domain specific Retrieval Augmented Generation (RAG) system designed to answer legal questions over diverse Indian Supreme Court judgments. Our suggested method utilises the rhetorical structure of court documents, in contrast to conventional  RAG pipelines that handle documents as plain, unstructured text. Every sentence in the dataset is annotated with one of the six rhetorical roles: Arguments, Fact, Statute, Precedent, RatioOfTheDecision, and RulingByPresentCourt. The pipeline includes document chunking, retrieval, and prompt generation and these annotations. The system is evaluated on a collection of 30 annotated Supreme Court judgments drawn from three legal domains: Civil, Corporate, and Criminal law — with 10 judgments per domain. The dataset comprises Supreme Court judgments from Indian Kanoon\footnote{\url{https://indiankanoon.org/}}, a public legal search engine. To support various legal NLP applications, the plain-text decisions were selected across various legal domains, ensuring a fair mix of subject matter and document lengths. Performance is measured using three automated RAG evaluation metrics: Faithfulness, Answer Relevancy, and Contextual Relevancy. Figure~\ref{fig:methodology} presents a block diagram  of the rhetorical-role-aware RAG pipeline. Corpus-level and  chunking details are shown in Table~\ref{tab:corpus_stats}.


\begin{table}[h]
\centering
\caption{Corpus and Chunking Statistics}
\label{tab:corpus_stats}
\begin{tabular}{lrrrr}
\hline
\textbf{Statistic} & \textbf{Civil} & \textbf{Corporate} & \textbf{Criminal} & \textbf{Total} \\
\hline
Number of judgments       & 10    & 10     & 10     & 30     \\
Total sentences           & 4381  & 1806   & 2915   & 9102   \\
Avg. sentences / judgment & 438.1 & 180.6  & 291.5  & 303.4  \\
Total chunks              & 557   & 220    & 345    & 1122   \\
Avg. chunks / judgment    & 55.7  & 22.0   & 34.5   & 37.4   \\
Avg. sentences / chunk    & 7.84  & 8.64   & 8.43   & 8.3    \\
\hline
\end{tabular}
\end{table}

\subsection{Rhetorical-Role-Aware Chunking}\label{subsec1}

Conventional fixed-size and recursive chunking strategies typically partition documents without considering the semantic function of individual sections. As a result, legally related information may be fragmented across multiple chunks, reducing the effectiveness of retrieval. To address this limitation, the proposed pipeline organizes document content according to rhetorical roles, producing semantically coherent chunks in which sentences serving similar legal purposes are grouped together. Sentence-level rhetorical role annotations were generated using the labeling model proposed in \citep{malik-etal-2022-semantic}. This model assigns each sentence to one of seven rhetorical categories:Argument, Statute, Fact, Ratio of the Decision, Precedent and Ruling by Lower Court. These categories represent the functional contribution of each sentence within a judicial decision, providing a structured view of the reasoning process and overall organization of legal judgments.

By integrating rhetorical roles into the chunking process, the retrieved context is better aligned with the intent of different legal queries, therefore, the retrieval component can obtain the legally relevant information which is also semantically meaningful. The chunking strategy is implemented through the following four-steps:
\begin{itemize}
    \item Role Grouping: Consecutive sentences labled with the same rhetorical role are combined to form a single semantic group.
    \item Small-Group Merging: Groups containing fewer than two sentences are merged with the immediately preceding group, reducing the number of isolated micro-chunks that do not contain useful information and may introduce noise into the retrieval index.
    \item Long-Group Sub-splitting: Groups containing more than fifteen sentences are divided into overlapping windows of fifteen sentences, with a two-sentence overlap between adjacent windows. This overlap preserves local contextual continuity ensuring the closely related legal information remain intact at chunk boundaries.
\end{itemize}

\begin{figure}[h]
  \centering
  \includegraphics[width=\linewidth]{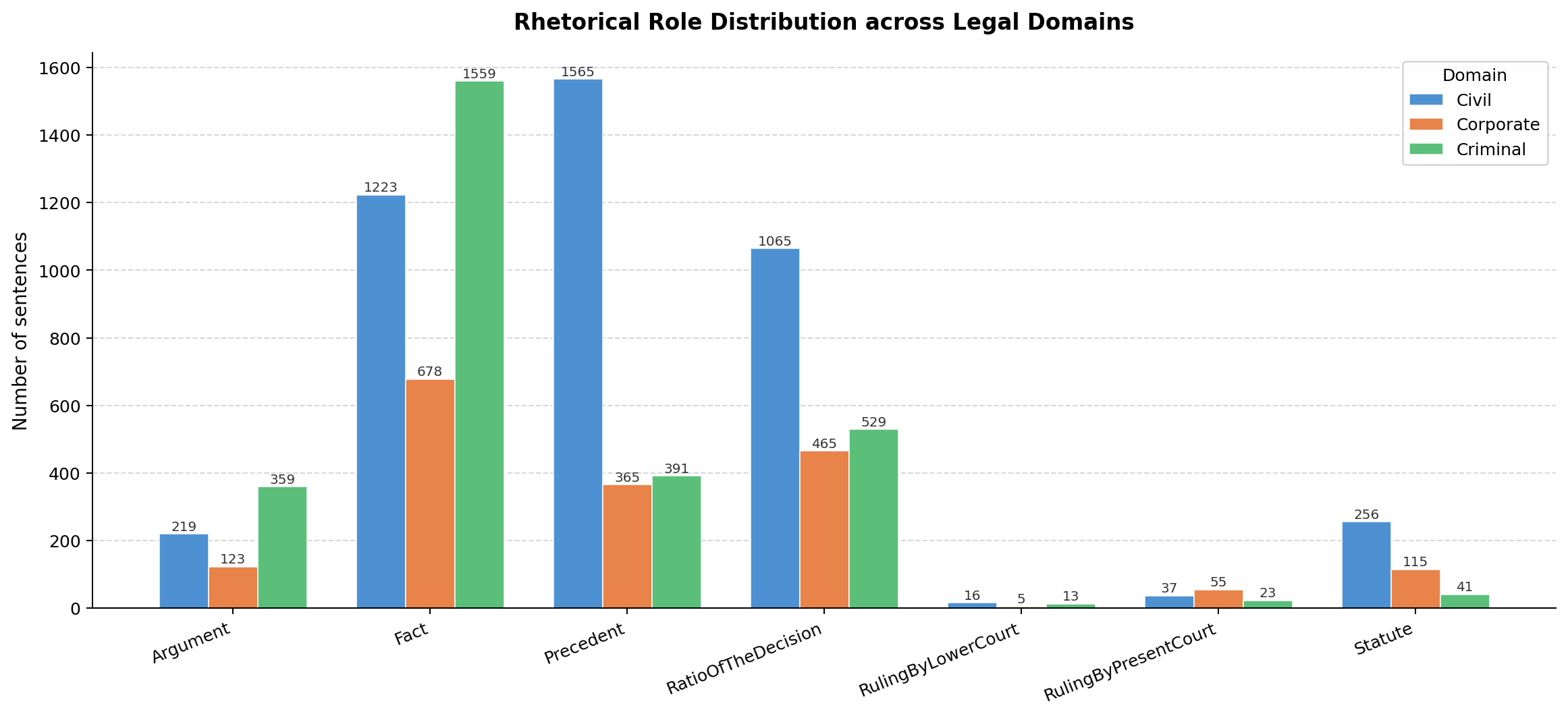}
  \caption{Rhetorical role distribution across judgments}
  \label{fig:roles}
\end{figure}

The distribution of rhetorical roles across domains is presented in 
Table~\ref{tab:role_distribution} and illustrated in Figure~\ref{fig:roles}. 
Fact and Precedent are the most frequently occurring roles across all 
three domains.

\begin{table}[h]
\centering
\caption{Rhetorical Role Distribution across Legal Domains (sentence counts)}
\label{tab:role_distribution}
\begin{tabular}{lrrrr}
\hline
\textbf{Role} & \textbf{Civil} & \textbf{Corporate} & \textbf{Criminal} & \textbf{Total} \\
\hline
Argument              & 219  & 123 & 359  & 701  \\
Fact                  & 1223 & 678 & 1559 & 3460 \\
Precedent             & 1565 & 365 & 391  & 2321 \\
RatioOfTheDecision    & 1065 & 465 & 529  & 2059 \\
RulingByLowerCourt    & 16   & 5   & 13   & 34   \\
RulingByPresentCourt  & 37   & 55  & 23   & 115  \\
Statute               & 256  & 115 & 41   & 412  \\
\hline
\textbf{Total}        & \textbf{4381} & \textbf{1806} & \textbf{2915} & \textbf{9102} \\
\hline
\end{tabular}
\end{table}

\subsection{Dual Index Construction}\label{subsec2}

To facilitate complementary retrieval strategies, the chunked corpus is indexed using two parallel representations: a dense semantic index and a sparse lexical index. The dense index helps to capture semantic relationships between legal texts, whereas the sparse index retains exact lexical information. By combining these two representations, the framework benefits from both semantic understanding and precise keyword matching during retrieval.


A lightweight bi-encoder trained for passage retrieval converts each chunk into a dense vector of fixed length. Instead of depending solely on keyword matching, the encoder is trained to learn semantic representations that enable the system to retrieve relevant legal passages even when the wording of the query differs from that of the document. User queries are encoded using the same model to ensure that both queries and document chunks are located in the same semantic embedding space. Before encoding, an instruction based retrieval prompt is added to each query to improve its alignment with the passage representations. For effective similarity searching, resultant embeddings for each document chunks are kept in the dense retrieval index.

Alongside the dense retrieval index, a sparse lexical index is also constructed using the BM25Okapi implementation of BM25 \citep{robertson2009probabilistic}. Each document chunk is split using simple whitespaces, and the resulting tokens are used to calculate the term-frequency statistics necessary for computing the BM25 score. The lexical retrieval complements dense retrieval by selecting passages that contain exact query key terms. This is especially helpful in the case of legal documents when the passage relevance is determined by mentions of case names, statutory citations, section numbers, or other legal terms. Integrating dense and sparse retrieval allows the framework to combine semantic understanding with keyword matching, improving its ability to retrieve both semantically similar and lexically relevant information.

\subsection{Retrieval Pipeline}\label{sec4}



\subsubsection{Query Intent Classification}
\begin{figure}[h]
  \centering
  \includegraphics[width=\linewidth]{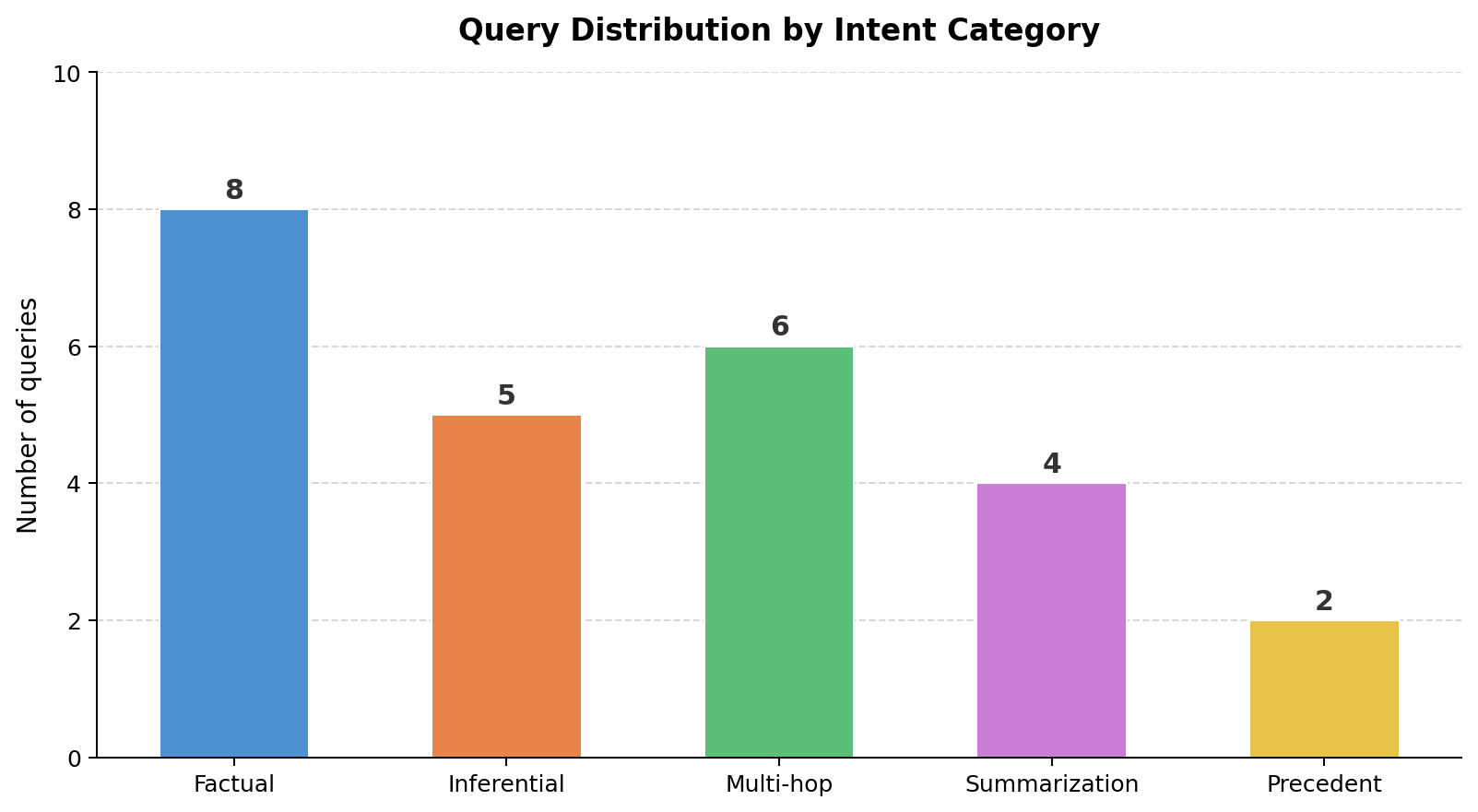}
  \caption{Query distribution by intent category}
  \label{fig:queries}
\end{figure}

The difference between various types of legal questions is that they all need different kinds of data in order to be answered correctly. While some seek only factual information, others involve logical reasoning in various parts of a judgment to arrive at the conclusion. To accommodate these differences, each query is alloted to one of five predefined intent categories. Factual queries form the largest group, with eight queries covering aspects such as party identification, bench composition, evidence, and the main issues of a case. The next largest group consists of six multi-hop queries, which require information to be gathered and integrated from different parts of the judgment. Inferential queries account for five instances and are intended to assess the court’s reasoning and interpretation of legal issues. The summarization category contains four queries that request an overview of the case. Precedent queries form the smallest group among all query types, comprising two queries that primarily focus on judicial precedents and statutory references. This distribution reflects the inherent diversity of legal information requirements, where factual and reasoning-based questions occur more often and in greater variety than queries focused mainly on legal citations. The queries are distributed across five intent categories as shown in 
Table~\ref{tab:intent_roles} and Table~\ref{tab:query_stats}, with the 
distribution visualised in Figure~\ref{fig:queries}.\\
\begin{table}[h]
\caption{Query intent categories and corresponding target rhetorical roles.}
\label{tab:intent_roles}
\begin{tabularx}{\linewidth}{@{}lXX@{}}
\toprule
\textbf{Intent} & \textbf{Description} & \textbf{Target Rhetorical Roles} \\
\midrule
Factual &
Direct and verifiable information explicitly stated in the judgment &
Fact \\

Inferential &
Court reasoning, legal interpretation, and justification of decisions &
Ratio Of The Decision, Argument \\

Multi-hop &
Questions requiring synthesis of information from multiple sections &
Fact, Ratio Of The Decision, Argument, Precedent \\

Summarization &
Overall case summary, procedural history, and outcome &
Ratio Of The Decision, Fact, Ruling By Present Court \\

Precedent &
References to prior cases, legal authorities, and statutory provisions &
Precedent, Statute \\
\bottomrule
\end{tabularx}
\footnotetext{Each query is assigned to a predefined intent category, which is subsequently mapped to the rhetorical roles most likely to contain relevant information.}
\end{table}

\begin{table}[h]
\centering
\caption{Query Distribution by Intent Category}
\label{tab:query_stats}
\begin{tabular}{lrrr}
\hline
\textbf{Intent} & \textbf{Query count} & \textbf{Percentage (\%)} & \textbf{Query-document pairs} \\
\hline
Factual       & 8  & 32.0  & 240 \\
Inferential   & 5  & 20.0  & 150 \\
Multi-hop     & 6  & 24.0  & 180 \\
Summarization & 4  & 16.0  & 120 \\
Precedent     & 2  & 8.0   & 60  \\
\hline
\textbf{Total} & \textbf{25} & \textbf{100.0} & \textbf{750} \\
\hline
\multicolumn{4}{l}{\small Query-document pairs = Query count $\times$ 30 judgments.} \\
\end{tabular}
\end{table}

\subsubsection{Role Aware Filtering}
To improve retrieval precision, candidate passages are filtered correspondoing to the rhetorical roles associated with the predicted query intent. Only segments from the roles that are most relevant to the query are kept Using the rhetorical structure of court rulings, this step guaranties that the retrieved evidence obtained aligns with the user's information requirements.

\subsubsection{Cross-Encoder Reranking}
The filtered candidate passages are reranked using a cross encoder model. In contrast to the retrieval stage, the cross encoder computes each query–passage pair together and assigns a score based on their combined context. As a result, the model can generate a more accurate ranking of retrieved data and capture semantic links.

Once the cross-encoder has produced the final set of passages, these are passed to a large language model, which generates the response. The input prompt fed to the model consists of five distinct elements: a system-level instruction that confines the model to use the provided context and the output is a predefined fallback response whenever the evidence does not contain an answer; a query specific instruction, which defines how the model should respond — for example, requiring exact factual information for factual queries or a step by step chain of reasoning for multi-hop queries; a limited sequence of recent conversational turns to maintain continuity across a multi-turn exchange; the retrieved passages themselves with their own rhetorical roles; and finally, the query is appended at the end of the prompt just before the model generates the response. By designing the prompt to the query's intent, the system is able to ensure that the reasoning style of the model meets the requirements of the different types of legal questions.

\section{Results and Discussions}\label{sec5}
\subsection{Metrics}

The system is evaluated using metrics commonly employed for Retrieval-Augmented Generation (RAG) systems. Each metric is computed separately by a large language model acting as an evaluation judge.

\begin{equation}
\text{Faithfulness} = \frac{|\text{Supported claims}|}{|\text{Total claims in answer}|}
\end{equation}

Answer Relevancy is the percentage of statements in the generated answer that are relevant to the input query:

\begin{equation}
\text{Answer Relevancy} = \frac{|\text{Relevant statements}|}{|\text{Total statements in answer}|}
\end{equation}

Contextual Relevancy measures the how much of the retrieved context contains information that is relevant to the input query

\begin{equation}
\text{Contextual Relevancy} = \frac{|\text{Relevant retrieved nodes}|}{|\text{Total retrieved nodes}|}
\end{equation}

Evaluation scores are on a scale of 0 to 1, where higher value corresponds better performance. Across all legal domains considered in our work, Gemini-2.5-Pro consistently achieves the highest Faithfulness scores, demonstrating that its responses remain closer to the facts found in the source than those of any  other models. In comparison, scores for the Answer Relevancy show only minor differences across models and legal domains. This consistency suggests that the proposed intent-aware prompting strategy works well in generating responses that remain aligned with the user's query, regardless of the language model.
Table~\ref{tab:civil} presents evaluation results on Civil law judgments. 
Table~\ref{tab:corporate} and Table~\ref{tab:criminal} present results 
for Corporate and Criminal law domains respectively.

\begin{table}[h]
\caption{Evaluation Results: Civil Law Domain}\label{tab:civil}
\begin{tabular}{@{}llll@{}}
\toprule
Model & Faithfulness & Answer Relevancy & Contextual Relevancy \\
\midrule
Qwen3-32B & 0.966 & 0.860 & 0.443 \\
Gemini-2.5-flash & 0.970 & 0.906 & 0.378 \\
Gemini-2.5-Pro & 0.973 & 0.901 & 0.376 \\
\bottomrule
\end{tabular}
\footnotetext{Note: Scores for civil documents are normalized on a scale of 0.0 to 1.0.}
\end{table}

\begin{table}[h]
\caption{Evaluation Results: Corporate Law Domain}\label{tab:corporate}
\begin{tabular}{@{}llll@{}}
\toprule
Model & Faithfulness & Answer Relevancy & Contextual Relevancy \\
\midrule
Qwen3-32B A & 0.958 & 0.923 & 0.389 \\
Gemini-2.5-flash & 0.973 & 0.885 & 0.395 \\
Gemini-2.5-Pro & 0.985 & 0.914 & 0.385 \\
\bottomrule
\end{tabular}
\footnotetext{Note: Evaluation focused on contract analysis and compliance documentation.}
\end{table}

\begin{table}[h]
\caption{Evaluation Results: Criminal Law Domain}\label{tab:criminal}
\begin{tabular}{@{}llll@{}}
\toprule
Model & Faithfulness & Answer Relevancy & Contextual Relevancy \\
\midrule
Qwen3-32B & 0.951 & 0.884 & 0.441 \\
Gemini-2.5-flash & 0.984 & 0.881 & 0.386 \\
Gemini-2.5-Pro & 0.991 & 0.903 & 0.377 \\
\bottomrule
\end{tabular}
\footnotetext{Note: Performance metrics reflect the complexity of case law and evidentiary standards.}
\end{table}

\section{Conclusion}\label{sec13}










This paper presented a rhetorical-role-aware Retrieval-Augmented Generation (RAG) framework for QA over Indian Supreme Court judgments. The proposed approach retrieves information that is not only relevant to the query but also comes from the most appropriate parts of a judgment. To improve the quality of information retrieved prior to response generation, the system integrates role-aware chunking, hybrid retrieval, intent-based filtering, and cross-encoder reranking.

\section{Limitations}
This evaluation was carried out solely on Indian Supreme Court judgments, and the scores reported come entirely from automated LLM-based metrics rather than review by legal experts. Because of this, the results may not carry over cleanly to other judicial systems, lower-court documents, or non-English judgments, and without human evaluation, it is harder to identify errors in legal reasoning that automated metrics might miss.

\bibliography{sn-bibliography}

\end{document}